\documentclass[12pt, a4paper]{article}
\usepackage{amsmath,amssymb,cite,comment,bm,url}

\input{colordvi.tex}

\usepackage{ifpdf}
\ifpdf        
\usepackage{graphicx, hyperref, xcolor}     %   usepackage without driver option
\else     % For (p)LaTeX + dvipdfmx
\usepackage[dvipdfmx]{graphicx, hyperref, xcolor}     %   usepackage with driver option
\fi

\usepackage[height=21.5cm,width=16.5cm]{geometry}
\definecolor{rossoferrari}{HTML}{D9073D}
\definecolor{mediumblue}{HTML}{0000CD}
\hypersetup{% hyperref option list
setpagesize=false,
bookmarksnumbered=true,
bookmarksopen=true,
colorlinks=true,
linkcolor=rossoferrari,
urlcolor=mediumblue,
citecolor=mediumblue,
}

\newcommand{\keV}{\mathrm{keV}}
\newcommand{\MeV}{\mathrm{MeV}}
\newcommand{\GeV}{\mathrm{GeV}}

\begin{document}
%%%%%%%%%%%%%%%%%%%%%%%%%%%%%%%%%%%%%%%%%%%%%%%%%%

%%%%%%%%%%%%%%%%%%%%%%%%%%%%%%%%%%%%%%%%%%%%%%%%%%
\begin{titlepage}

\begin{center}

\hfill 

\vskip 2cm

{\Large \bf 
Gravitational Production of Hidden Photon \\[.3em] Dark Matter in light of the XENON1T Excess
}

\vskip 1cm

{\large
Kazunori Nakayama$^{(a,b)}$ and Yong Tang$^{(c,d,e)}$
}

\vskip 1cm
$^{(a)}${\em Department of Physics, Faculty of Science,\\
The University of Tokyo,  Bunkyo-ku, Tokyo 113-0033, Japan}\\[.3em]
$^{(b)}${\em Kavli IPMU (WPI), The University of Tokyo,  Kashiwa, Chiba 277-8583, Japan}\\[.3em]
$^{(c)}${\em School of Astronomy and Space Sciences, \\ University of Chinese Academy of Sciences (UCAS), Beijing, China}\\[.3em]
$^{(d)}${\em National Astronomical Observatories, Chinese Academy of Sciences, Beijing, China}\\[.3em]
$^{(e)}${\em School of Fundamental Physics and Mathematical Sciences, \\
	Hangzhou Institute for Advanced Study, UCAS, Hangzhou 310024, China }
\end{center}
\vskip 1cm

\begin{abstract}

Recently, the XENON1T experiment has reported an excess in the electronic recoil events. The excess is  consistent with the interpretation of absorption of 3\,keV bosonic dark matter, for example, hidden photon dark matter with kinetic mixing of the order of $10^{-15}$. We point out that the minimally gravitational production provides a viable mechanism for obtaining a correct relic hidden photon abundance. We present parameter dependence of the hidden photon dark matter abundance on the inflationary scale $H_{\rm inf}$ and also the reheating temperature $T_{\rm R}$. We show that the inflationary Hubble scale and reheating temperature are both bounded from below, $H_{\textrm{inf}}\gtrsim 7\times 10^{11}\,\textrm{GeV},\; T_{\rm R}\gtrsim 10^2\,\textrm{GeV}$. In particular, the high-scale inflation is consistent with 3\,keV hidden photon dark matter.

\end{abstract}

\end{titlepage}

%\tableofcontents

\renewcommand{\thepage}{\arabic{page}}
\setcounter{page}{1}
\renewcommand{\thefootnote}{\#\arabic{footnote}}
\setcounter{footnote}{0}
%%%%%%%%%%%%%%%%%%%%%%%%%%%%%%%%%%%%%%%%%%%%%%%%%%

\newpage

Recently, the XENON1T collaboration has reported excess events in the electronic recoil with the recoil energy around $2$--$7$\,keV~\cite{Aprile:2020tmw}. The excess may be interpreted as a contribution from the solar axion~\cite{Aprile:2020tmw}, but this interpretation is inconsistent with the stellar cooling constraint, in particular the observation of white dwarfs and red giants~\cite{Giannotti:2017hny}.
On the other hand, various connections of the XENON1T excess with particle physics models, constraints and implications have been investigated in~\cite{Takahashi:2020bpq, Kannike:2020agf, Alonso-Alvarez:2020cdv, Fornal:2020npv,Boehm:2020ltd, 1802403,1802395, 1802391,1802374, 1802394, 1802396, DiLuzio:2020jjp, Du:2020ybt, Su:2020zny,Bally:2020yid, Harigaya:2020ckz}. Absorption of bosonic dark matter (DM) may also give similar signals~\cite{Pospelov:2008jk, Arisaka:2012pb}. 
One of the good candidates is the hidden photon DM with mass $m_V \simeq 3$\,keV and the kinetic mixing parameter $\epsilon \sim 10^{-15}$, which is also shown to be consistent with the anomalous cooling of horizontal branch stars~\cite{Alonso-Alvarez:2020cdv}. 
%Fig.~\ref{fig:signal} shows the signal event shape along with the XENON1T data points for the hidden photon mass $m_V= 2.7$\,keV and the kinetic mixing parameter $\epsilon=7\times 10^{-16}$ (see Eq.~(\ref{S}) for definition), which apparently shows that the XENON1T excess events can be well fitted by the hidden photon DM. The indicated hidden photon mass $m_V \simeq 3$\,keV and the kinetic mixing parameter $\epsilon \sim 10^{-15}$ is also shown to be consistent with the anomalous cooling of horizontal branch stars~\cite{Alonso-Alvarez:2020cdv}. 

The viable production of keV DM is not trivial since the constraints from astrophysical observations are severe for keV DM produced from thermal plasma. For example, the recent Lyman-$\alpha$ gives the lower bound~$\gtrsim 5.3$~keV~\cite{Palanque-Delabrouille:2019iyz} if it is thermally produced. Hence, a keV DM candidate would require other viable production mechanism. In this short note, we focus on the hidden photon DM interpretation and show that the gravitational production mechanism is consistent with such a $\sim 3$\,keV hidden photon DM with explicit parameter dependence on the inflationary energy scale and the reheating temperature.

%%%%%%%%%%%%%%%%
\begin{figure}
	\centering
	\begin{tabular}{cc}
		\includegraphics[width=0.65\columnwidth]{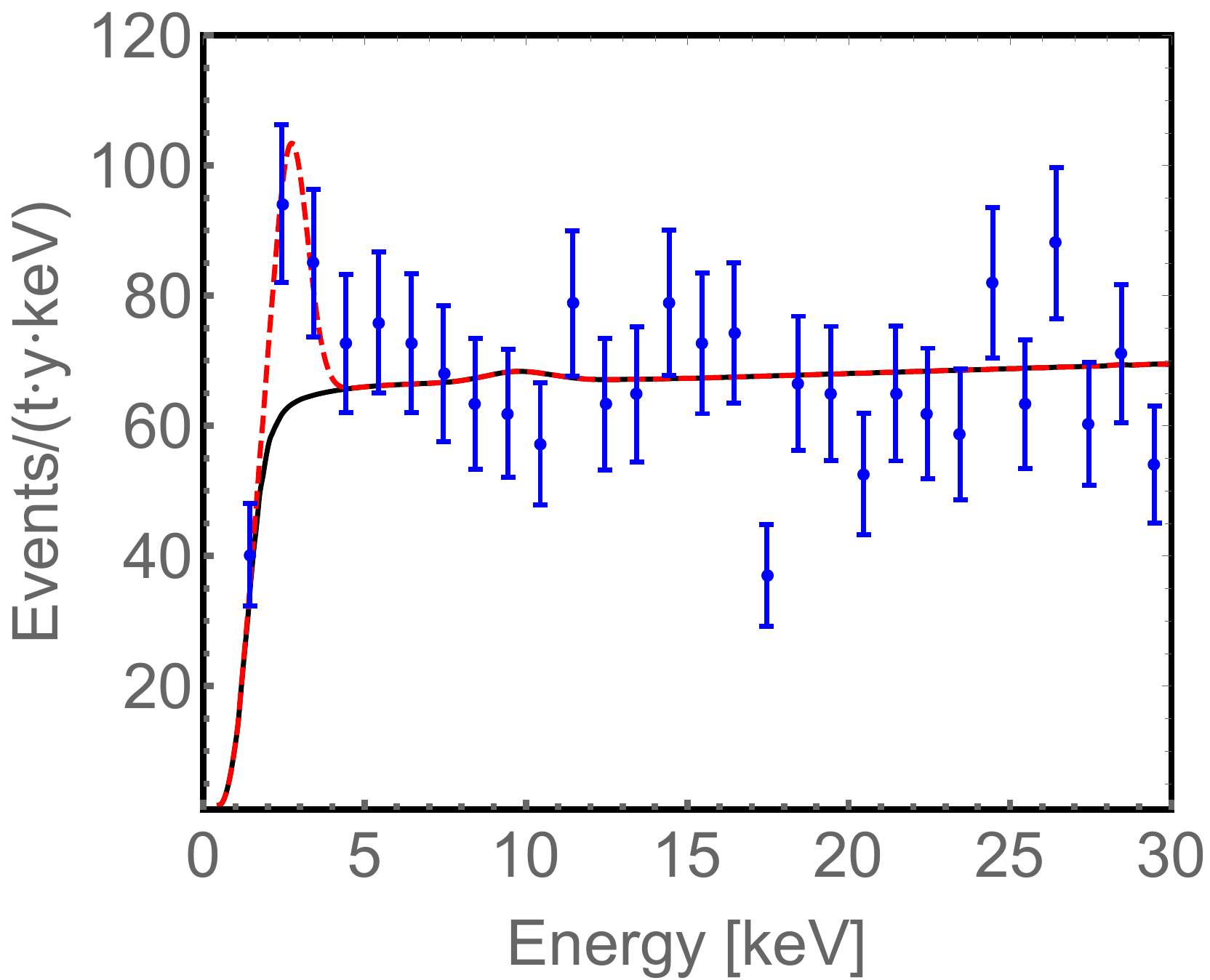}
	\end{tabular}
	\caption{The signal shape for hidden photon with $m_V = 2.7$\,keV and $\epsilon=7\times 10^{-16}$. The blue dots are the events observed by XENON1T~\cite{Aprile:2020tmw}, the black solid curve is the background, and the red dashed curve displays the events including the absorption of hidden photon.}
	\label{fig:signal}
\end{figure}
%%%%%%%%%%%%%%%%

The most relevant action of the hidden photon $V_\mu$ and the Standard Model (SM) electromagnetic photon $A_\mu$ for our discussions is given by
\begin{align}
	S &= \int d^4 x\sqrt{-g}\left( -\frac{1}{4}V_{\mu\nu}V^{\mu\nu} -\frac{1}{4}F_{\mu\nu}F^{\mu\nu} - \frac{1}{2}m_V^2 V_\mu V^\mu +\frac{\epsilon}{2}V_{\mu\nu}F^{\mu\nu} \right) \label{S} \\
	&= \int d^4 x\sqrt{-g}\left( -\frac{1}{4}\overline V_{\mu\nu}\overline V^{\mu\nu} -\frac{1}{4}\overline F_{\mu\nu}\overline F^{\mu\nu} - \frac{1}{2}\overline m_V^2 \overline V_\mu \overline V^\mu\right),   \label{S_bar}
\end{align}
where $V_{\mu\nu}\equiv \partial_\mu V_\nu-\partial_\nu V_\mu$ and $F_{\mu\nu}\equiv \partial_\mu A_\nu-\partial_\nu A_\mu$ are the field strength tensor of the hidden photon and SM photon, respectively. In the second line we have defined $\overline A_\mu=A_\mu-\epsilon V_\mu$, $\overline V_\mu = \sqrt{1-\epsilon^2} V_\mu$ and $\overline m_V^2=m_V^2/\sqrt{1-\epsilon^2}$.\footnote{
	If we start from the hypercharge photon, instead of the electromagnetic photon, and introduces the kinetic mixing of $\epsilon_Y$, we end up with (\ref{S}) after the electroweak symmetry breaking with $\epsilon = \epsilon_Y \cos\theta_W$ where $\theta_W$ is the Weinberg angle.
} In the second basis, the kinetic terms are diagonalized while the SM particles are coupled with the hidden photon with coupling suppressed by $\epsilon$. The mass term is understood as a result of the Stuckelberg mechanism or the Higgs mechanism with the Higgs excitation being assumed to be heavy enough so that its dynamics is safely ignored.
The massive hidden photon with kinetic mixing has rich phenomenological implications~\cite{Jaeckel:2010ni,Fabbrichesi:2020wbt}.

Due to the kinetic mixing, electrons in atoms can absorb keV hidden photon DM~\cite{Pospelov:2008jk, Arisaka:2012pb}, similar to photoelectric effect, and register as the electronic recoil events in XENON1T. The rate for such absorption is given by
\begin{equation}
R\simeq \frac{1.7\times 10^{29}}{A}\epsilon^2 \left(\frac{\keV}{m_V}\right)\left(\frac{\sigma_{\textrm{pe}}}{\textrm{barn}}\right)\textrm{ton}^{-1}\textrm{year}^{-1},
\end{equation}
with $A\approx 131$ the atomic mass of Xenon and $\sigma_{\textrm{pe}}$ the cross section of photoelectric effect~\cite{pecross}. Here we have used the local dark matter density $\rho \simeq 0.3\GeV/$cm$^3$. For $m_V=2.7~\keV$ and $\epsilon=7\times 10^{-16}$, we find the number of absorption events is $\sim 40$ for XENON1T's exposure of $0.65$ tonne-years. 
The energy spectrum of these events is sharply located at the mass of hidden photon but smeared by the finite energy resolution of the detection. The resolution~\cite{Aprile:2020yad} is parametrized by Gaussian distribution with uncertainty $\sigma$,
\begin{equation}
	\frac{\sigma}{E} =\left(\frac{31.71}{\sqrt{E}}+0.15\right)\%.
\end{equation}
For the reconstructed energy at $E\simeq 2.7~\keV$, the relative resolution is about $19.45\%$. In Fig.~\ref{fig:signal}, we show the spectrum of signal events (red dashed curve) on top of the backgrounds (black solid), in contrast to the observed events (blue dots). We can see the spectral shape is consistent with the excess at $2-3~\keV$.

Next, we discuss how such hidden photons were produced in the early universe. As mentioned above, for such a light hidden photon to be DM, thermal production is disfavored by Lyman-$\alpha$ observation. Therefore, some viable production mechanisms are required. 
Conversion of thermal SM photons to the hidden photon does not lead to enough amount of relic hidden photon for the parameters we are interested in~\cite{Redondo:2008ec,An:2013yfc,Redondo:2013lna}.
There are several proposed mechanisms so far: tachyonic instability due to the axionic scalar coupling to the hidden photon~\cite{Agrawal:2018vin,Co:2018lka,Bastero-Gil:2018uel}, production from the dark Higgs dynamics~\cite{Dror:2018pdh}, from the cosmic strings associated with the spontaneous breaking of hidden U(1) symmetry~\cite{Long:2019lwl}, vector coherent oscillation~\cite{Nelson:2011sf,Arias:2012az,AlonsoAlvarez:2019cgw,Nakayama:2019rhg,Nakayama:2020rka,Nakai:2020cfw} and the gravitational production~\cite{Graham:2015rva,Ema:2019yrd,Ahmed:2020fhc}.
Among them we focus on the gravitational production mechanism since it is ubiquitous: such a contribution is unavoidable as far as we consider the inflationary universe and actually it is enough to reproduce the DM abundance for $m_V \sim \mathcal O(1)$\,keV, as shown below.

The gravitational production of hidden photon DM was first discussed in Ref.~\cite{Graham:2015rva} in the case of $m_V \ll H_{\rm inf}$ with the assumption of instant reheating, where $H_{\rm inf}$ denotes the Hubble scale during inflation.
In Ref.~\cite{Ema:2019yrd} it was extended to the case of delayed reheating $H_{\rm R} \ll H_{\rm inf}$ and also the case of heavy hidden photon: $m_V \gtrsim H_{\rm inf}$, where $H_{\rm R}$ is the Hubble scale at the completion of reheating.
%As far as the kinetic mixing parameter is much smaller than unity, the calculation of the gravitational production rate remains intact. 

The essential procedures to investigate the gravitational production of hidden photon are the following. First, we shall solve the equation of motion for $V_\mu (\overline{V}_\mu) $ in the Friedman-Robertson-Walker background, whose metric is given by 
\begin{align}
	ds^2=-dt^2 + a^2(t)\delta_{ij}dx^idx^j  = a^2(\tau)\left(-d\tau^2 + \delta_{ij}\right)dx^idx^j,
\end{align}
with $a$ the scale factor and $\tau$ the conformal time. Since the kinetic mixing is so tiny in our interested case, for production we do not distinguish $V$ and $\overline V$ below. 
%In the Appendix we will discuss small effect from nonzero $\epsilon$. 
It is usually more convenient to work with Fourier mode by decompose the vector field into the transverse and longitudinal mode: 
\begin{align}
	V_\mu(\vec x,\tau) =  \int \frac{d^3k}{(2\pi)^3} V_\mu(\vec k,\tau) e^{i\vec k\cdot\vec x}; ~~~~~~
	\vec V(\vec k,\tau) = \vec V_{\rm T}(\vec k,\tau) +\frac{\vec k}{|\vec k|} V_{\rm L}(\vec k,\tau),
\end{align}
where the transverse mode satisfies $\vec k\cdot \vec V_{\rm T}=0$. Then the action $S$ in eq.~(\ref{S}) is the sum of the transverse part,
\begin{align}
	S_{\rm T}&= \int \frac{d^3kd\tau}{(2\pi)^3}\frac{1}{2}\left(|\vec V_{\rm T}'|^2 -(k^2+a^2m_V^2) |\vec V_{\rm T}|^2 \right),
\end{align}
and the longitudinal one,
\begin{align}
	S_{\rm L}&= \int \frac{d^3kd\tau}{(2\pi)^3}\frac{1}{2}\left( \frac{a^2 m_V^2}{k^2+a^2m_V^2}|V_{\rm L}'|^2 -a^2m_V^2 |V_{\rm L}|^2 \right) \\
	&=  \int \frac{d^3kd\tau}{(2\pi)^3}\frac{1}{2}\left\{ |\widetilde V_{\rm L}'|^2 -\left[k^2+a^2m_V^2-\frac{k^2}{k^2+a^2m_V^2}\left(\frac{a''}{a}-\frac{a^{\prime 2}}{a^2}\frac{3a^2 m_V^2}{k^2+a^2m_V^2} \right) \right] |\widetilde V_{\rm L}|^2 \right\}.
\end{align}
We have defined $\widetilde V_{\rm L}\equiv \sqrt{a^2m_V^2/(k^2+a^2m_V^2)} V_{\rm L}$ and the prime denotes the derivative with respect to $\tau$. Both $V_T$ and $\widetilde{V}_L$ would satisfy the following equation of motion, 
\begin{align}
	&V''+\omega_{\textrm{eff}}^2 V=0,\\
	&\omega_{\textrm{eff}}^2=\begin{cases}
	\displaystyle k^2+a^2m_V^2 & \textrm{for } V=V_T\\
	\displaystyle k^2+a^2m_V^2-\frac{k^2}{k^2+a^2m_V^2}\left(\frac{a''}{a}-\frac{a^{\prime 2}}{a^2}\frac{3a^2 m_V^2}{k^2+a^2m_V^2} \right) & \textrm{for } V=\widetilde{V}_L
\end{cases}.
\end{align}
Finally, after solving $V$, we can substitute in the energy-momentum tensor $T_{\mu\nu}=\cfrac{-2}{\sqrt{-g}} \cfrac{\delta S}{\delta g^{\mu \nu}}$ and evaluate the energy density $\rho_{\rm HP}=T_{00}$, see Ref.~\cite{Ema:2019yrd} for details.

The hidden photon abundance from the gravitational production, in terms of the energy density-to-entropy density ratio, is given by~\cite{Ema:2019yrd}
\begin{align}
	\frac{\rho_{\rm HP}}{s}\simeq \begin{cases}
		\displaystyle \frac{3}{2048\pi}\frac{m_V T_{\rm R}H_{\rm inf}}{M_{\rm Pl}^2} & {\rm for}~~H_{\rm inf} < m_V\\
		\displaystyle \frac{1}{32\pi^2}\frac{T_{\rm R}H_{\rm inf}^2}{M_{\rm Pl}^2} & {\rm for}~~H_{\rm R}< m_V < H_{\rm inf} \\
		\displaystyle \left( \frac{90}{\pi^2 g_*} \right)^{1/4}\frac{1}{32\pi^2}\frac{m_V^{1/2} H_{\rm inf}^2}{M_{\rm Pl}^{3/2}} & {\rm for}~~m_V <H_{\rm R}
	\end{cases},
	\label{rhoHP}
\end{align}
where $M_{\rm Pl}$ is the reduced Planck scale and we have defined the reheating temperature $T_{\rm R}$ through $T_{\rm R}=(90/\pi^2g_*)^{1/4}\sqrt{H_{\rm R}M_{\rm Pl}}$. As usual, the universe is assumed to be matter-dominated during the reheating since the inflaton harmonic oscillation behaves as non-relativistic matter. Practically the case of $H_{\rm inf} < m_V$ is irrelevant since it predicts too small hidden photon abundance and it cannot be DM for $m_V\sim \mathcal O(1)$\,keV. 
Below we briefly discuss how to understand (\ref{rhoHP}) for $m_V<H_{\rm inf}~$ intuitively.

The transverse mode production is negligible compared with the longitudinal one and we focus on the latter.
The longitudinal action reduces to the same action as a minimal scalar for $k> am_V$ and hence the quantum fluctuation in inflationary epoch results in $\left<|\widetilde V_{\rm L}|^2\right> \simeq a_{\rm end}^2H_{\rm inf}^2/(2k^3)(2\pi)^3\delta(k-k')$ for the superhorizon modes $a_{\rm end} H_{\rm inf} < k < a_{\rm end} m_V$ at the end of inflation, where $a_{\rm end}$ denotes the scale factor at the end of inflation. Now we want to estimate the abundance when the Hubble parameter $H$ becomes equal to the hidden photon mass: $H=m_V$. 
It is found that the high frequency mode $k > k_*\equiv a_*m_V$ is redshifted away rather rapidly, where $a_*$ denotes the scale factor at $H=m_V$, while the low frequency mode $k<k_*$ has an initially suppressed energy density already at $a=a_{\rm end}$. The dominant contribution to the final energy density comes from the mode $k\sim k_*$.
Thus the energy density at $H=m_V$ is evaluated as
\begin{align}
	\rho_{\rm HP}(a_*)\simeq \frac{1}{2}\left( \frac{k_*}{a_*} \right)^2\left( \frac{H_{\rm inf}}{2\pi} \right)^2 = \frac{m_V^2 H_{\rm inf}^2}{8\pi^2}.
\end{align}
When $H<m_V$, the energy density scale as $\propto a^{-3}$ as an ordinary non-relativistic matter. It leads to the second and third line of (\ref{rhoHP}).\footnote{
	This expression of the hidden photon energy density is similar to the scalar coherent oscillation energy density as if the initial amplitude of the scalar field is $H_{\rm inf}/(2\pi)$.
}
Numerically, we have the present hidden photon relic density for $m_V<H_{\rm R}$ as
\begin{equation}
\rho_{\rm HP} \sim \rho_{\textrm{DM}}\times \sqrt{\frac{m_V}{3\,\textrm{keV}}}\times \left(\frac{H_{\rm inf}}{7\times 10^{11}\,\textrm{GeV}}\right)^2,
\end{equation}
where $\rho_{\textrm{DM}}$ is the average energy density of DM at present. When the reheating temperature after inflation is low, i.e. $H_{\rm R}<m_V<H_{\textrm{inf}}$, we have 
\begin{equation}
\rho_{\rm HP} \sim \rho_{\textrm{DM}}\times \frac{T_{\rm R}}{10^6\,\textrm{GeV}} \times \left(\frac{H_{\rm inf}}{7\times 10^{11}\,\textrm{GeV}}\right)^2,
\end{equation}
which is independent of the mass $m_V$. 

\begin{figure}
\centering
\begin{tabular}{cc}
\includegraphics[width=0.7\columnwidth]{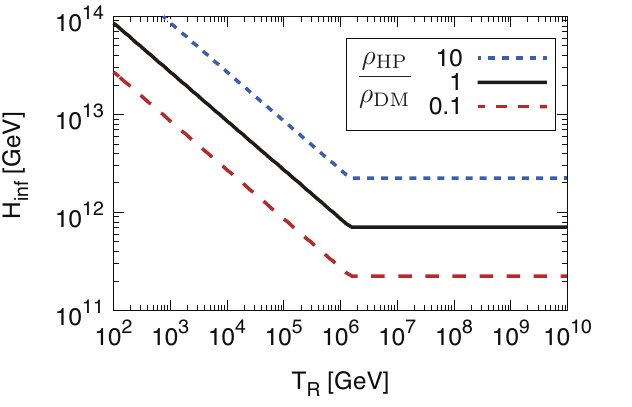}
\end{tabular}
\caption{The illustration of the energy density of dark photon $\rho_{\rm HP}$ for $m_V=3$\,keV. The solid line shows how the correct relic abundance would require the proper values of inflation scale $H_{\textrm{inf}}$ and reheating temperature $T_{\rm R}$. The dotted and dashed lines correspond to ten times larger and smaller, respectively. The turning points indicate $H_{\rm R} \simeq m_V$. }
\label{fig:relic}
\end{figure}

In Fig.~\ref{fig:relic}, we illustrate how the relic density of hidden photon depends on the reheating temperature $T_{\rm R}$ and inflation scale $H_{\textrm{inf}}$, when fixing the mass of hidden photon as $m_V=3$\,keV. The solid line gives the right relic abundance of DM, while the dotted and dashed lines correspond to ten times larger and one tenth smaller, respectively. In the large $T_{\rm R}$, the production is independent of $T_{\rm R}$, which shows that even the instant reheating is valid. When $T_{\rm R}$ is small enough such that $H_{\rm R}$ is smaller than the hidden photon's mass, we would need large inflation scale $H_\textrm{inf}$ to compensate the production loss because the relic density now depends on $T_{\rm R}$ linearly. The turning points of these curves occur around $H_{\rm R}\simeq m_V$. 
Since the inflation scale $H_{\textrm{inf}}$ is bounded from above, $H_{\textrm{inf}}\lesssim 10^{14}$\,GeV, which is based on the non-observation of primordial gravitational waves by the {\it Planck} satellite~\cite{Akrami:2018odb}, the correct relic abundance of hidden photon would give the lower bound on the Hubble scale and reheating temperature,
\begin{equation}
H_{\textrm{inf}}\gtrsim 7\times 10^{11}\,\textrm{GeV},~~~~~ T_{\rm R}\gtrsim 10^2\,\textrm{GeV} .
\end{equation}
The reheating temperature is also bounded from above, $T_{\rm R}\lesssim 10^{15}$\,GeV, corresponding to the instant reheating limit. 

%%%%%%%%%%%%%%%%
%%%%%%%%%%%%%%%%

Interestingly, the delayed reheating scenario ($H_{\rm R}\ll H_{\rm inf}$) opens up a possibility for high-scale inflation to be consistent with hidden photon DM. For maximally possible inflation scale $H_{\textrm{inf}}\sim 10^{14}$\,GeV, the reheating temperature is predicted to be around the weak scale: $T_{\rm R}\sim 10^2$\,GeV.
In this case, it is possible to probe the primordial gravitational waves through the observation of the B-mode polarization in the cosmic microwave background anisotropy. On the other hand, it is below the sensitivity of future space-based direct gravitational wave detectors due to too low $T_{\rm R}$~\cite{Nakayama:2008ip,Nakayama:2008wy,Kuroyanagi:2008ye}.

One of the good aspects of the hidden photon DM with gravitational production is that it is not constrained from the limit on the DM isocurvature fluctuation. In contrast to light scalar DM, the isocurvature fluctuation spectrum is strongly blue for the hidden photon DM~\cite{Graham:2015rva} and hence practically there is no effect on the cosmic microwave background anisotropy on the cosmological scale. Thus even the high-scale inflation does not suffer from the isocurvature constraint.
Another aspect is that it is an unavoidable contribution since the gravity is universal. In this sense it gives lower bound on the hidden photon abundance. It is possible that hidden photon has interactions with other sector and the production is more efficient, which, however, is highly model-dependent. 
For example, the inflaton may decay into hidden photon pair if there is a coupling between them. Assuming that hidden photons are non-relativistic in the present universe, such a contribution can be evaluated as
\begin{align}
	\left( \frac{\rho_{\rm HP}}{s}\right)_{\phi\to2V}&\simeq {\rm Br}_{\phi\to2V}\frac{3T_{\rm R}}{2}\frac{m_V}{m_\phi}\\
	&\sim 5\times 10^{-9}\,{\rm GeV}\times {\rm Br}_{\phi\to2V} \left( \frac{T_{\rm R}}{10^{10}\,{\rm GeV}} \right)
	\left( \frac{m_V}{3\,{\rm keV}} \right)\left( \frac{10^{13}\,{\rm GeV}}{m_\phi} \right),
\end{align}
where $m_\phi$ denotes the inflaton mass and $ {\rm Br}_{\phi\to2V}$ denotes the branching ratio of the inflaton into two hidden photons. It can be much smaller than the DM abundance if the branching ratio is much smaller than unity. 
%Note that the hidden photons produced in this way are often relativistic even in the present universe, depending on the choice of $m_\phi$ and $T_{\rm R}$. 
Note that the hidden photons produced in this way could be relativistic even in the present universe if $m_\phi/T_{\rm R} \gg 10^6$.
In such a case the hidden photon may be regarded as dark radiation and the requirement is just ${\rm Br}_{\phi\to2V} \lesssim \mathcal O(0.1)$ to avoid constraint on the effective number of extra neutrino species. Actually the inflaton decay to the hidden photon can easily be suppressed if the inflaton is charged under some (approximate) symmetry so that the coupling like $\phi V_{\mu\nu}V^{\mu\nu}$ is forbidden.

Because of the kinetic mixing, there is also a thermal contribution due to scattering, for instance, $e^- + \gamma \rightarrow e^- + V_\mu$. The production rate for such a process is given by $\Gamma \sim \epsilon^2 \alpha^2 T$, where $\alpha\simeq 1/137$ is the fine-structure constant and $T$ is the temperature of thermal plasma. In comparison to Hubble parameter in the radiation-dominant era, $H\sim T^2/M_{\rm Pl}$, we observe $\Gamma/H\propto \epsilon^2 \alpha^2 M_{\rm Pl} /T$, inversely proportional to the temperature. This indicates that such a production is most effective at low temperature, namely at $T\sim\MeV$ before $e^{\pm}$ decouple. We can estimate the number ratio of hidden photon to ordinary photon is about $10^{-13}$ for $\epsilon\sim 10^{-15}$. For keV hidden photon, such a thermal contribution only constitutes $10^{-10}$ of the total energy density at present, therefore it can be safely neglected, without conflicting any observational constraints. See Refs.~\cite{Redondo:2008ec,An:2013yfc,Redondo:2013lna} for more detailed discussion.

We also note that the kinetic mixing does not affect the gravitational production rate. One may worry about the breaking of conformal invariance due to the kinetic mixing and the SM gauge coupling through the trace anomaly~\cite{Dolgov:1993vg,Prokopec:2001nc}, which might result in the gravitational production of the hidden photon. To study it, it is convenient to go to the interaction basis:
\begin{align}
	S &= \int d^4 x\sqrt{-g}\left[ -\frac{1}{4}\mathcal V_{\mu\nu}\mathcal V^{\mu\nu} -\frac{1}{4}\mathcal F_{\mu\nu}\mathcal F^{\mu\nu} - \frac{1}{2}m_V^2 \left(\mathcal V_\mu+\frac{\epsilon}{\sqrt{1-\epsilon^2}}\mathcal A_\mu \right)^2 +e \mathcal A_\mu J^\mu \right], \label{S_int}
\end{align}
where $\mathcal V_\mu\equiv V_\mu-\epsilon A_\mu$ and $\mathcal A_\mu \equiv \sqrt{1-\epsilon^2}A_\mu$ and $J^\mu$ represents the SM U(1) current. It is evident that the SM photon and hidden photon are completely decoupled in the limit $m_V\to 0$ and hence the kinetic mixing leads to neither the breaking of conformal invariance nor the particle production in the massless limit. Thus the effect of kinetic mixing on the gravitational production rate is suppressed by powers of $\epsilon$ compared with the tree level effect studied in this letter.

The kinetic mixing can also induce the decay of hidden photon into three photons~\cite{Pospelov:2008jk}, $V_\mu \rightarrow 3\gamma$, through the box diagram with $e^\pm$ running in the loop. The decay width of $V_\mu \rightarrow 3\gamma$ is estimated as
\begin{equation}
\Gamma_{V\rightarrow 3\gamma}\sim 1.4\times 10^{-16}\epsilon^2\frac{m^9_V}{m^8_e}.
\end{equation}
The keV photons from the decay would be observed as X-ray in space telescope. Requiring the decay width is smaller than $10^{-27}$s${^{-1}}$, we can put a constraint on $m_V\epsilon^{2/9}\lesssim 0.1\,\keV $, which is satisfied in our interested parameter space. If the kinetic mixing is between hidden photon and hypercharge gauge field, there would be also mixing between $V_\mu$ and neutral weak gauge boson $Z_\mu$, which induces the decay of $V$ into two neutrinos. However, this mixing is further suppressed by the mass ratio, $m^2_V/m^2_Z$. Therefore, the decay width would be 
\begin{equation}
	\Gamma_{V\rightarrow \nu \bar{\nu}}\sim 10^{-2} \epsilon^2 \frac{m^5_V}{m_Z^4}.
\end{equation}
Since it is very difficult to observe  $\keV$ low-energy neutrinos at current experiments, as long as this decay width is small than $10^{-17}$s${^{-1}}$, $V$'s lifetime would be longer than the age of universe. This puts an upper limit on $m_V\epsilon^{1/2}\lesssim \keV,$ which is much weaker than the one from $V_\mu \rightarrow 3\gamma$. 

Before we summarize, we comment on the possible effects if this hidden photon is associated with Higgs mechanism. The immediate effect is that we then also need to consider the gravitational production of the hidden Higgs boson, which is investigated in Ref.~\cite{Ema:2018ucl}. How such a hidden Higgs boson can affect the relic abundance of hidden photon is model-dependent. To simplify the discussion, we can consider two limiting cases: small hidden gauge coupling ($g_H$) with large hidden Higgs vacuum expectation value ($v_H$) and large $g_H$ with small $v_H$. In both cases we need $g_H v_H \sim {\rm keV}$ to reproduce the hidden photon DM mass.
In the former case, there would be no effect and our discussions above are unaffected since $v_H$ can take very large value (say, $10^{15}$\,GeV) and the hidden Higgs dynamics can be completely ignored. In the latter case, we would need to solve the coupled equations of motion for hidden Higgs and photon. Several things can happen: gravitational (and other) production of the hidden Higgs boson and restoration of symmetry during/after inflation followed by the formation of cosmic strings, which would significantly affect the hidden photon abundance~\cite{Dror:2018pdh,Long:2019lwl}.
Dedicated investigation of such a scenario would be beyond our scope here. 

In summary, hidden photon DM with $m_V\sim 3\,$keV and $\epsilon\sim 10^{-15}$ is a good candidate to explain the XENON1T excess in electronic recoil events and satisfies the relevant observational constraints. The gravitational production works well for such a mass region for reasonable inflationary scale and the wide range of the reheating temperature. In particular, taking into account of the effect of delayed reheating, the high-scale inflation can be consistent with the hidden photon DM scenario.

%%%%%%%%%%%%%%%%%%%%%%%%%%%%%%%%%%%%%%%%%%%%
\section*{Acknowledgments}
%%%%%%%%%%%%%%%%%%%%%%%%%%%%%%%%%%%%%%%%%%%%

This work was supported by the Grant-in-Aid for Scientific Research C (No.18K03609 [KN]) and Innovative Areas (No.17H06359 [KN]).
This work was supported by Natural Science Foundation of China under Grant No.~11851302 and by the Fundamental Research Funds for the Central Universities [YT].

%%%%%%%%%%%%%%%%%%%%%%%%%%%%%%%%%%%%%%%%%%%%%%%%%%

%%%%%%%%%%%%%%%%%%%%%%%%%%%%%%%%%%%%%%%%%%%%%%%%%%

%%%%%%%%%%%%%%%%%%%%%%%%%%%%%%%%%%%%%%%%%%%%%%%%%%
\end{document}